\documentclass[aps,prb,reprint,longbibliography,citeautoscript,floatfix]{revtex4-2}
\usepackage[utf8]{inputenc}
\usepackage[T1]{fontenc}

\usepackage{graphicx} %

\usepackage{hyperref}
\usepackage{xcolor}
\usepackage{amsmath}
\usepackage{siunitx}
\DeclareSIUnit\angstrom{\text{\AA}}
\usepackage{dcolumn}
\usepackage{xfrac}
\usepackage{microtype}
\usepackage{tikz}

\usepackage{graphicx}

\usepackage{CJK}

\newcommand{\mytitle}{Faceting transition in aluminum as a grain boundary phase transition}

\hypersetup{
  final,
  pdfborder={0 0 0},
  pdftitle={\mytitle},
  pdfauthor={Choi, Brink},
  colorlinks=true,
  urlcolor=blue,
  linkcolor=blue,
  citecolor=blue
}

\begin{document}
\frenchspacing

\title{\mytitle}

\begin{CJK*}{UTF8}{mj}

\author{Yoonji Choi (최윤지)}
\affiliation{Max Planck Institute for Sustainable Materials,
  Max-Planck-Stra\ss{}e 1, 40237 D\"usseldorf, Germany}

\author{Tobias Brink}
\email{t.brink@mpie.de}
\affiliation{Max Planck Institute for Sustainable Materials,
  Max-Planck-Stra\ss{}e 1, 40237 D\"usseldorf, Germany}

\date{4 August 2025}

\begin{abstract}
  Grain boundaries facet due to anisotropic grain boundary energies:
  While the faceted boundary has a larger area than the corresponding
  straight boundary, a significantly lower energy of the facets
  compared to a straight segment can drive the faceting. This picture
  is complicated by faceting/defaceting transitions where the free
  energy difference between the two states depends on the temperature.
  Such transitions have been observed and modeled before, but their
  exact nature was not fully understood. Here, we use atomistic
  computer simulations to show that a well known faceting transition
  in $\Sigma 3$ $[11\overline{1}]$ tilt grain boundaries in Al is in
  fact a grain boundary phase transition (also called complexion
  transition). This means that the faceted and defaceted boundaries
  are associated with different atomic structures, which have
  different thermodynamic stability ranges. At low temperatures, the
  grain boundary phase associated with faceting is stable, while at
  high temperatures the flat grain boundary phase is stable. We also
  report on our thorough tests of Al interatomic potentials for this
  purpose, which include comparisons to density-functional theory
  calculations. Our chosen potential performs well for our grain
  boundaries. As a consequence we were able to obtain results that
  align with previous experimental results.
\end{abstract}

\maketitle

\end{CJK*}

\newcounter{supplfigctr}
\renewcommand{\thesupplfigctr}{S\arabic{supplfigctr}}
\newcounter{suppltabctr}
\renewcommand{\thesuppltabctr}{S-\Roman{suppltabctr}}
{\refstepcounter{supplfigctr}\label{fig:suppl:grip-clustering-011}}
{\refstepcounter{supplfigctr}\label{fig:suppl:grip-clustering-112}}
{\refstepcounter{supplfigctr}\label{fig:suppl:zippers}}
{\refstepcounter{supplfigctr}\label{fig:suppl:free-energy-class-qm}}
{\refstepcounter{supplfigctr}\label{fig:suppl:latconst}}
{\refstepcounter{supplfigctr}\label{fig:suppl:dft-fcc}}
{\refstepcounter{supplfigctr}\label{fig:suppl:dft-B1}}
{\refstepcounter{supplfigctr}\label{fig:suppl:dft-B2}}
{\refstepcounter{supplfigctr}\label{fig:suppl:dft-square}}
{\refstepcounter{supplfigctr}\label{fig:suppl:grip-allpots}}
{\refstepcounter{supplfigctr}\label{fig:suppl:derivatives-a}}
{\refstepcounter{supplfigctr}\label{fig:suppl:derivatives-b}}

\section{Introduction}

\begin{tikzpicture}[remember picture,overlay]
  \node [anchor=north west, font=\footnotesize, align=left,
         text width=7.05in, xshift=0.75in, yshift=0.75in, inner xsep=0pt]
        at (current page.south west)
        {Published in:\\
         \href{https://doi.org/10.1103/2dnf-zdz8}
              {Choi and Brink, Phys.~Rev.~Mater.~\textbf{9}, 083607 (2025)}
         \hfill
         DOI: \href{https://doi.org/10.1103/2dnf-zdz8}
                   {10.1103/2dnf-zdz8}};
\end{tikzpicture}%
Grain boundaries (GBs) are abundant planar defects in polycrystalline
materials. While these defects are not thermodynamically stable, the
transformation from polycrystals to single crystals is often so slow
at the application temperature that the defective system is in a deep
local free energy minimum \cite{Gottstein2004, Lejcek2010}. In
well-relaxed systems, GBs therefore still transform into energetically
favorable states \cite{Priester2013}. If the excess free energy of a
GB is highly anisotropic with regards to the crystallographic plane it
occupies, GB faceting can occur \cite{Wagner1974, Brokman1981,
  Hsieh1989, Straumal2001, Hamilton2003, Wu2009, Banadaki2016}. There,
the GB increases its GB area by splitting into facets on low-energy
planes, thereby decreasing its overall free energy despite its higher
GB area. Typically, the facets have a driving force to grow in size,
because the line defects separating the facets (facet junctions) are
also connected to an energy cost \cite{Dimitrakopulos1997, Pond1997,
  Hamilton2003, Wu2009, Medlin2017, Hadian2018, Brink2024}.

Reversible faceting/defaceting transitions---wherein the GB switches
between a faceted and a flat state with changing temperature---have
been observed. The most famous example is the $\Sigma 3$
$[11\overline{1}]$ (011) tilt GB in Al, which splits into \{112\}
facets at low temperatures \cite{Hsieh1989, Wu2009, Straumal2016}. The
explanation for this reversible defect transition must lie in the
relative free energies of the faceted and flat GBs. The free energy of
a GB is, however, not just a function of the macroscopic GB parameters
(misorientation between the crystallites and GB plane), but also of
the GB structure. In analogy to bulk phases, we can treat these
different structures of the interface as GB phases \cite{Frolov2015a},
also called complexions \cite{Tang2006, Dillon2007, Cantwell2014,
  Cantwell2020}. These are distinct from bulk phases that nucleate at
the GB, such as precipitates or wetting phases \cite{Kaplan2013,
  Cantwell2020}, but are instead constrained to the interface and can
be described by interface thermodynamics \cite{GibbsVol1, Hart1968,
  Hart1972, Cahn1982, Rottman1988, Frolov2012, Frolov2012a,
  Kaplan2013, Cantwell2014, Frolov2015a, Cantwell2020}. We can
categorize GB phase transitions into congruent transitions, where the
macroscopic GB parameters remain the same while the atomic structure
of the GB changes; and non-congruent transitions, where also the
macroscopic GB parameters change. Faceting/defaceting transitions
would thus be non-congruent GB phase transitions.

In $[11\overline{1}$] tilt GBs of fcc metals with misorientation
$\theta < \ang{60}$ (the $\Sigma 3$ boundary has a misorientation of
exactly \ang{60}), several GB phases were found
[Fig.~\ref{fig:bicryst}(d)] \cite{Meiners2020, Meiners2020a,
  Langenohl2022, Langenohl2023, Brink2023, Saood2024}. On the
near-(011) plane, congruent transitions between the ``domino'' and
``pearl'' phases were found in Cu \cite{Meiners2020, Langenohl2022}
and predicted by simulations for many fcc metals \cite{Brink2023}. On
the near-(112) planes, only a ``zipper'' phase was found
\cite{Meiners2020a, Langenohl2023}. Recently, the domino phase was
identified as consisting of nanoscale zipper facets, which is only
possible for $\theta < \ang{60}$ \cite{Brink2024}. For
$\theta = \ang{60}$, these facets become macrofacets. Thus, if a
congruent domino-to-pearl GB phase transition exists, this raises the
question if the faceting/defaceting transition in $\Sigma3$
$[11\overline{1}]$ (011) GBs is a non-congruent version of this
transition. If a pearl-like phase exists, it might be more stable at
high temperatures than a faceted zipper GB. Here, we explore this
hypothesis with atomistic simulatons.

\section{Theory}

\subsection{Grain boundary phases and their bicrystallography}

\begin{figure*}
    \centering
    \includegraphics[]{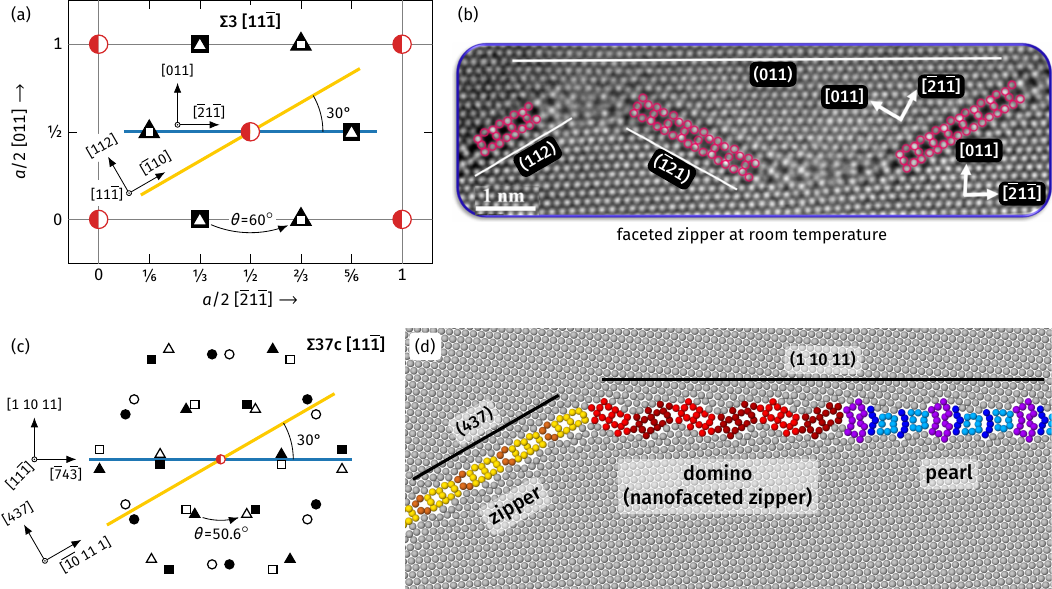}
    \caption{(a) Dichromatic pattern of $\Sigma 3$ $[11\overline{1}]$
      tilt GBs. Black and white symbols represent the atoms of the two
      crystallites, while the shape of the symbols represents the
      different $(11\overline{1})$ planes on which these atoms
      lie. The black and white symbols coincide in the projection, but
      are located on different planes. The red data points are
      coincidence sites. The crystal directions refer to the
      coordinate system of the lower crystallite.  Two possible
      (quasi-)symmetric GB planes are indicated. (b) At room
      temperature, faceted zipper structures were observed
      experimentally in Al on \{112\} planes. The structure on the
      (011) plane is not reported in the literature. Image from
      Ref.~\cite{Saood2024} with crystallographic directions and
      planes adjusted to the convention of the present paper. Reused
      under the terms of the Creative Commons Attribution 4.0
      International license
      (\url{https://creativecommons.org/licenses/by/4.0/}).  (c)
      Excerpt of the dichromatic pattern of a $\Sigma 37$c
      $[11\overline{1}]$ tilt GB. The red data point is a coincidence
      site. The full dichromatic pattern can for example be found in
      the supplemental material of Ref.~\cite{Langenohl2022}.  (b)
      Snapshot of a simulation of the corresponding GB phases. The
      domino phase consists of nanoscale facets of zipper structures
      \cite{Brink2024}. This nanofaceting is impossible in the
      $\Sigma3$ GB, where the facets are energetically driven to grow
      \cite{Brink2024}. We hypothesize that pearl-like phases may
      exist on the (011) plane of $\Sigma3$.}
    \label{fig:bicryst}
\end{figure*}

In order to better understand the GB phases involved, we will quickly
discuss the crystallography and atomic structures of the relevant
GBs. There are two sets of high-symmetry GB planes in $\Sigma3$
$[11\overline{1}]$ tilt GBs ($\theta = \ang{60}$), as indicated in the
dichromatic pattern in Fig.~\ref{fig:bicryst}(a). These planes are
inclined by \ang{30} towards each other. The structure of the \{112\}
GB planes has been identified before as the zipper GB phase
[Fig.~\ref{fig:bicryst}(b)] in Al at room temperature \cite{Saood2023,
  Saood2024}. This GB has a mirror symmetry across its GB plane and is
therefore a symmetric GB. The $\Sigma3$ $[11\overline{1}]$ \{011\} GBs
have crystallographically equivalent planes on both sides of the GB,
but the GB plane is not a mirror plane. Those GBs are called
quasi-symmetric in the nomenclature of Morawiec
\cite{Morawiec2012}. At room temperature, faceting into \{112\} zipper
facets was observed in Al \cite{Saood2024}, but if a non-faceted GB
phase existed on the \{011\} planes, it could be responsible for a
non-congruent faceting/defaceting transition.

A related transition does indeed exist for $\theta <
\ang{60}$. Figure~\ref{fig:bicryst}(c)--(d) shows the $\Sigma37$c
$[11\overline{1}]$ tilt GB as an example. We similarly obtain a set of
symmetric \{347\} GB planes and quasi-symmetric \{1~10~11\} GB planes,
inclined by \ang{30} towards each other. While the $\Sigma37$c
$[11\overline{1}]$ \{347\} GBs also exhibit zipper GB phases, the
\{1~10~11\} planes are home to the domino and pearl phases. A
congruent domino-to-pearl transition exists \cite{Meiners2020,
  Langenohl2022, Brink2023}, but a non-congruent faceting transition
does not: any \{1~10~11\} plane with faceted zipper structures
exhibits an unusual attractive interaction between the facet junctions
and leads to nanofaceting \cite{Brink2024}. This nanofaceted phase is
the domino phase, as can also be confirmed by visual inspection of the
motifs in Fig.~\ref{fig:bicryst}(d). In the present work we look for a
pearl-like phase in $\Sigma3$ GBs. Since nanofaceting is not possible
for $\Sigma3$ boundaries \cite{Brink2024}, the congruent
domino-to-pearl transition would become a non-congruent transition
from zipper facets to a flat pearl-like phase.

\subsection{Grain boundary thermodynamics}

GB phases can be described with few modifications from standard
thermodynamics \cite{GibbsVol1, Hart1968, Hart1972, Cahn1982,
  Rottman1988, Frolov2012, Frolov2012a, Frolov2015a}. Most
importantly, GBs do not exist in thermodynamic equilibrium and we will
thus have to work with metastable states in which the existence of a
GB is postulated. As a further simplification for the purpose of this
paper, we assume that the bulk regions do not undergo phase
transitions, but only the GB does. Additionally, we only consider
pure, single-component materials. The Gibbs free energy $G$ of a
bicrystal can be separated into \cite{Frolov2015a}
\begin{equation}
  \label{eq:gibbs-sep}
  G = G_1 + G_2 + \gamma A + G_\text{surface},
\end{equation}
where $G_1$ and $G_2$ are the free energies of the two crystallites,
$G_\text{surface}$ accounts for the surface, $\gamma$ is the GB free
energy and $A$ is the GB area. Ignoring surface and bulk phase
transitions, we can describe GB phase transitions fully via excess
properties. We assume
$G_1/N_1 = G_2/N_2 = G_\text{bulk}/N_\text{bulk}$, where $N$ is the
number of atoms and $G_\text{bulk}$ is the Gibbs free energy of a
defect-free bulk region. Thus, we define the excess of any extensive
property $Z$ as \cite{Cahn1979, Frolov2012a}
\begin{equation}
  \label{eq:excess}
  [Z] = \frac{Z_\text{GB} - \frac{N_\text{GB}}{N_\text{bulk}}Z_\text{bulk}}{A},
\end{equation}
where $Z_\text{GB}$ is the value of $Z$ in a region containing a GB of
area $A$ with $N_\text{GB}$ atoms.

Important excess properties in the present paper are thus
\cite{Cahn1979, Frolov2012a}
\begin{equation}
  \label{eq:gamma}
  \gamma = [G] = [U] - T[S] - \sigma_{33}[V], %
\end{equation}
with internal energy $U$, entropy $S$, stress tensor $\sigma_{ij}$,
and volume $V$. Here, we use the boundary conditions from
Ref.~\cite{Frolov2012a}, where the $z$ direction (index 3, normal to
the GB plane) is under isobaric conditions with applied stress
$\sigma_{33}$. The $x$ and $y$ directions are fixed so that the bulk
lattice is unstrained at the given temperature $T$. In the present
case, we are not interested in systems under stress (except for
$\sigma_{33}$ in Appendix~\ref{sec:app:eval-pot}), so we use shear
stresses $\sigma_{31} = \sigma_{32} = 0$ and shear strain
$\varepsilon_{12} = 0$. The treatment of nonzero shear is described
elsewhere \cite{Frolov2012a}.

We assume ambient pressure for the faceting/defacting transition and
thus simplify most of our investigation to $\sigma_{3i} \approx 0$:
\begin{equation}
  \label{eq:gamma-simple}
  \gamma = [U] - T[S].
\end{equation}
For statics simulations, we also define the useful property
\begin{equation}
  \label{eq:gamma0}
  \gamma_0 = [E_\text{pot}]
\end{equation}
at $T = 0$, where $E_\text{pot}$ is the potential energy of the system
with all atoms in minimum-energy positions. In general,
quantum-mechanical zero-point vibrations mean that
$\gamma_0 \neq \gamma(T{=}0)$. In MD simulations, classical systems
are considered and $\gamma_0 = \gamma_\text{MD}(T{=}0)$.

Due to our definition (Eq.~\ref{eq:excess}) of the excess, $[N] = 0$
even if atoms are added or removed in the GB region. Nevertheless, the
difference of atoms per crystallographic plane between bulk and GB is
physically meaningful \cite{Frolov2013, Hickman2017}. Since individual
crystallographic planes may not be identifiable in the GB, the GB's
planar fraction can be defined as \cite{Frolov2013}
\begin{equation}
  \label{eq:excess-n}
  \hat{n} = \frac{N_\text{GB}}{N_\text{plane}} \mod 1,
\end{equation}
where $N_\text{plane}$ is the number of atoms in a defect-free
crystallographic plane with the same area $A$ as the GB. It is
necessary to vary $\hat{n}$ to find all GB phases
\cite{Frolov2013}. The value of $\hat{n}$ does not affect the
thermodynamics, but the transition between GB phases with different
$\hat{n}$ requires diffusion and is thus typically much slower
\cite{Frolov2013}.

To aid in classification of the GB phases we also used the GB excess
stresses, which are defined at $\sigma_{3i} = 0$ as
\begin{equation}
  \label{eq:excess-stress}
  \hat{\tau}_{ij} = \frac{\hat{\sigma}_{ij} V_\text{GB}}{A}
  \qquad \text{for $i,j \in \{1,2\}$,}
\end{equation}
where $\hat{\sigma}_{ij}$ is the residual stress in a region of volume
$V_\text{GB}$ that contains a GB with area $A$ \cite{Frolov2012a}. These
stresses are the result of the mismatch between bulk and GB while
fixing the strains $\varepsilon_{11}$, $\varepsilon_{22}$, and
$\varepsilon_{12}$.

\section{Methods}

Most calculations were performed with an empirical interatomic
potential, namely the embedded atom method (EAM) potential by
Zhakhovskii et al.\ \cite{Zhakhovskii2009} as downloaded from the NIST
Interatomic Potentials Repository \cite{ipr} (designated as ``Zha09''
in this paper). Other potentials were also tested as described in
Appendix~\ref{sec:app:eval-pot}. We used \textsc{lammps}
\cite{Thompson2022, Plimpton1995} to perform molecular dynamics (MD)
and statics simulations. In order to evaluate the potentials, we
conducted additional density-functional theory (DFT) simulations using
\textsc{vasp} 5.4.4 \cite{Kresse1993, Kresse1994, Kresse1996,
  Kresse1996a}. The simulations were set up, managed, and analyzed
with \textsc{pyiron} \cite{Janssen2019}.

\subsection{Grain boundary structure search}

The possible GB phases and their structures in $\Sigma$3
$[11\overline{1}]$ tilt GBs with (quasi-)symmetric GB planes (011) and
(112) were explored with the \textsc{grip} software
\cite{Chen2024}. This code automatically constructs GB supercells of
varying size (here we used variations from $1\times1$ to $5\times5$ in
the GB plane), applies a random displacement to one of the
crystallites, removes a random fraction of atoms from the GB region,
runs a canonical MD simulation with random temperature and duration
with 95\% probability, and then minimizes the atomic position with
regards to the potential energy. This is repeated many times to
collect various GB structures. High-energy structures are discarded
regularly. For the (011) GB plane, we also ran a structure search with
supercell sizes ranging from $1\times5$ to $1\times25$ (i.e., no
repetition along the tilt axis and 5 to 25 repetitions along $y$ in
the GB plane) to find additional faceted GBs with longer facet
lengths.

All simulations cells were periodic in the $x$ direction, which lies
along the tilt axis, and the $y$ direction, which also lies in the GB
plane. We used open surfaces in the $z$ direction normal to the GB.

\subsection{Temperature-dependent free energy calculation and
  molecular dynamics simulation of grain boundary phase transitions}

In single-element systems, the entropy is solely a vibrational
entropy. We therefore computed the GB excess free energy $\gamma$ as a
function of temperature $T$ for several GB phases of interest using
the quasi-harmonic approximation (QHA) \cite{Foiles1994,
  Freitas2018}. Here, force constant matrices were obtained with the
\texttt{dynamical\textunderscore{}matrix} command in
\textsc{lammps}. Then the Helmholtz free energies were calculated as
\begin{align}
  F_\text{classical} &= E_\text{pot} + k_BT\sum_{i=1}^{3N-3}\ln\frac{h\nu_i}{k_BT}\\
  F_\text{qm}        &= E_\text{pot} + k_BT\sum_{i=1}^{3N-3}\ln\left[2
                       \sinh \left(\frac{\frac{1}{2}h\nu_i}{k_BT}\right)\right],
\end{align}
where the sum is over the phonon eigenfrequencies $\nu_i$ (excluding
the three null values) obtained from the force constant matrix. We
calculated free energies under the assumption of classical oscillators
($F_\text{classical}$), correspoding to classical MD simulations, and
for quantum-mechanical oscillators ($F_\text{qm}$). These free
energies were obtained at different cell volumes corresponding to
different lattice constants. For each temperature, the relevant
lattice constant was chosen in one of two ways. For the classical
mechanics case, the lattice constant was obtained with an MD
simulation. For the quantum mechanics case, we used the minimum free
energy of the fcc structure as a function of volume at each
temperature. The latter approach can be unreliable for certain
potentials (see Appendix~\ref{sec:app:eval-pot}), which is why we
generally preferred the classical approach here. Finally, using a
defect-free reference, we obtain the GB free energy as
$\gamma(T) = [F(T)]$ as in Eq.~\ref{eq:gamma-simple}.

In addition to the predictions by free-energy calculations, we also
performed MD simulations on various GBs. We used a time integration
step of \SI{2}{fs}, Nos\'e--Hoover thermostats for the desired
temperature, and a barostat at zero pressure for the periodic
directions. These simulations were run for \SI{30}{ns} to observe the
resulting GB phase (faceted or straight). Visualization and structure
analysis using the polyhedral template matching (PTM) method
\cite{Larsen2016} were performed with \textsc{ovito}
\cite{Stukowski2010}.

\subsection{Density-functional theory}
\label{sec:methods:dft}

Due to the wide variety of available interatomic potentials for Al, we
had to decide on the most well-suited one for our GB calculations. We
therefore also calculated GB energy and excess volumes using
density-functional theory (DFT) on the GB structures with (011) planes
obtained with our structure search. To avoid vacuum in the DFT cell,
we produced fully periodic cells with two equivalent GBs. We tested
different amounts of bulk material between the GBs, up to cell sizes
of roughly 100 atoms. GB excess properties were converged at that
point (see Figs.~\ref*{fig:suppl:dft-B1}--\ref*{fig:suppl:dft-square}
in the Supplemental Material \cite{suppl}). The results for the (112) plane
(zipper structure) were taken from Refs.~\cite{Saood2023,Saood2023zenodo}.

We used the projector-augmented wave (PAW) method \cite{Bloechl1994}
within the generalized gradient approximation (GGA) with the
Perdew--Burke--Ernzerhof (PBE) parametrization \cite{Perdew1996}. We
employed PAW potentials \cite{Kresse1999} with three valence electrons
(3s$^2$ 3p$^1$). Following earlier work of one of us \cite{Saood2023},
high-accuracy parameters were used to ensure correct GB excess
properties, which are very sensitive to numerical errors. Based on
this, we used the previously obtained lattice constant of
\SI{4.040(5)}{\angstrom} for Al and performed simulations on the GBs
with a plane-wave energy cutoff of \SI{450}{eV} and a
$27 \times 27 \times 2$ $k$-point mesh on a $\Gamma$-centered
Monkhorst--Pack grid \cite{Monkhorst1976}. During structure
minimization, forces were converged to \SI{0.01}{eV/\angstrom}.

The cell sizes in $x$ and $y$ direction (GB plane) were fixed in
accordance with the fcc lattice constant to
$7.00 \times \SI{4.95}{\angstrom^2}$ ($1 \times 1$ supercell). We
varied the length of the cell in $z$ (normal to the GB) and fitted a
third-order polynomial to the energy--length curve. This allowed us to
obtain the energies and volumes as a function of $\sigma_{33}$, which
is the derivative of this curve normalized by the GB area. By
comparing a cell containing a GB with a similarly-sized defect-free
cell, we obtained the excess properites $[U]$ and $[V]$ as functions of
$\sigma_{33}$.

\section{Grain boundary phases of $\Sigma3$ [111] tilt boundaries}

 \begin{figure*}
    \centering
    \includegraphics{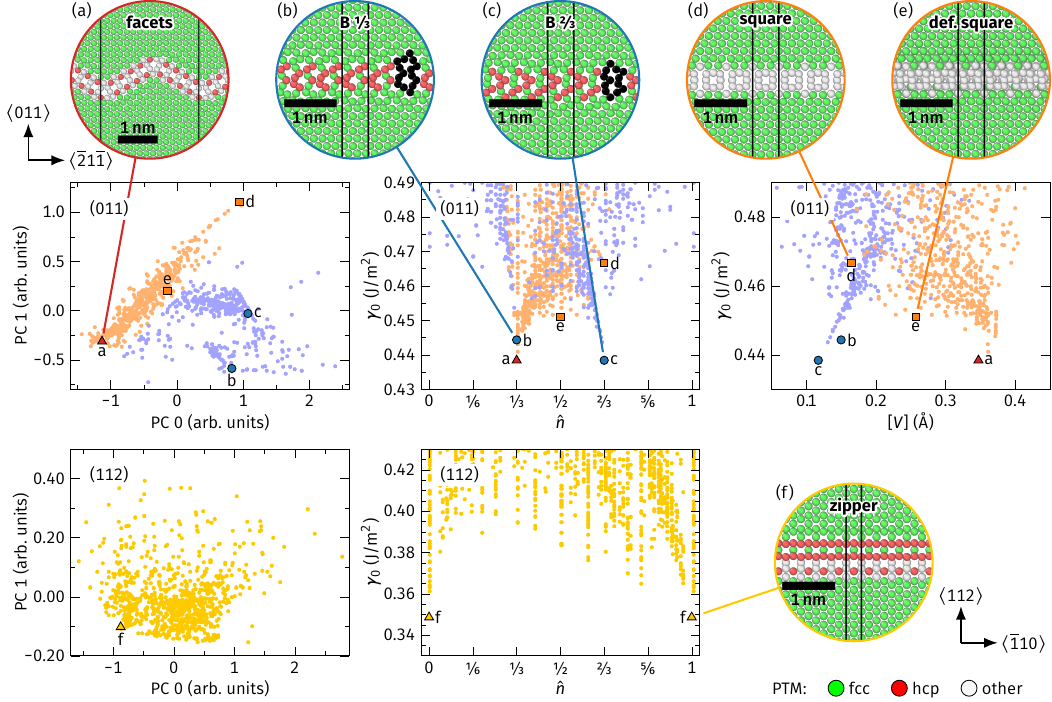}
    \caption{Results of the GB structure search. The graphs in the top
      row show excess properties obtained for the (011) GB plane, the
      bottom row shows excess properties for the (112) GB plane. For
      (011), the structures shown in subfigures (a)--(e) are
      representatives of low-energy GB phases. They are also indicated
      in the graphs. Using PCA, we plot the first two principal
      components PC 0 and PC 1. At least two groups, containing the
      square and faceted structures (orange), as well as the B
      structures (blue) can be distinguished. For (112) GB planes,
      only the zipper structure (f) and defective variants were
      found. Additional analysis and data is provided in
      Supplemental~Figs.~\ref*{fig:suppl:grip-clustering-011} and
      \ref*{fig:suppl:grip-clustering-112}. We show that all
      low-energy, faceted structures that we obtained are made of
      zipper facets in Supplemental Fig.~\ref*{fig:suppl:zippers}.}
    \label{fig:GRIP-zhakhovskii}
\end{figure*}

As a first step towards understanding the faceting transition in
$\Sigma3$ $[11\overline{1}]$ tilt boundaries, we performed an
extensive GB structure search with \textsc{grip} \cite{Chen2024}. We
performed this search for both of the possible (quasi-)symmetric GB
plane types, here we used the (011) and (112) planes
[Fig.~\ref{fig:bicryst}(a)]. The GB structure discovery yielded
thousands of structures with various GB energies, most of which are
defective local minima or high-energy structures. We excluded GBs with
$\gamma_0 > \SI{0.49}{J/m^2}$ for (011) and
$\gamma_0 > \SI{0.44}{J/m^2}$ for (112) from further analysis, since
they are too energetically unfavorable to occur in reality or in MD
simulations.

To categorize the remaining structures, we evaluated their excess
properties, the most important of which are plotted in
Fig.~\ref{fig:GRIP-zhakhovskii}. A principal component analysis (PCA)
\cite{Pearson1901, Hotelling1933, Hotelling1933a} performed on the
excess properties $\gamma_0$, $[V]$, $\hat{\tau}_{11}$, and
$\hat{\tau}_{22}$, reveals that the (011) GBs can be separated into at
least 2 clusters, while the (112) GBs all belong to one cluster (see
Supplemental~Figs.~\ref*{fig:suppl:grip-clustering-011} and
\ref*{fig:suppl:grip-clustering-112} for more details on this
analysis). By examination of the structures of low-energy GBs with
$(011)$ GB planes in each cluster, we identified two ``B'' structures
(with $\hat{n} = 1/3$ and 2/3), ``square'' structures, and faceted
structures. The B~\sfrac{1}{3} and square structures are structural
units that also occur in the pearl phase in $\Sigma37$c
$[11\overline{1}]$ $(1~10~11)$ GBs shown in Fig.~\ref{fig:bicryst}(d)
\cite{Langenohl2022}. The square motif also occurs in pearl GBs with
lower misorientation angles than the $\Sigma37$c GB \cite{Meiners2020,
  Brink2023}. On the (112) GB plane, we found only the zipper
structure \cite{Meiners2020a, Langenohl2023, Saood2023,
  Saood2024}. This zipper structure is also present in the faceted
structures on the (011) GB plane. While the structure of the short
facets in Fig.~\ref{fig:GRIP-zhakhovskii}(a) looks visually distinct
from the zipper structure in Fig.~\ref{fig:GRIP-zhakhovskii}(f), a
detailed analysis (Supplemental Fig.~\ref*{fig:suppl:zippers}) shows
that all faceted (011) boundaries always consist of zipper structures,
although they might be elastically distorted.

Both B~\sfrac{1}{3} and B~\sfrac{2}{3} are very similar in terms of
structure, GB energy, excess volume, and excess stresses. They are
thus most likely microstates of the same GB phase. The square motifs
without defects have a relatively high GB energy, which is why we also
show the lowest-energy, defective square-like motif at
$\hat{n} = 0.5$.  We conclude that three GB phases can occur in
$\Sigma3$ $[11\overline{1}]$ (011) tilt GBs in Al: B, square, and
faceted zipper. In a previous work on $[11\overline{1}]$ tilt GBs with
misorientation angles $\theta < \ang{60}$, it was found that the
zipper facets are nanofaceted, occurring as a domino GB phase
\cite{Brink2024}, while for $\theta = \ang{60}$ ($\Sigma3$) the facets
grow. In Cu, domino/pearl phase transitions were observed
\cite{Meiners2020, Langenohl2022}. This leads us to the hypothesis
that the faceting/defaceting transition in $\Sigma3$ GBs is analogous
to the domino/pearl transition, except for the length of the facets.

\section{Faceting transition as a grain boundary phase transition}

\subsection{Prediction for infinitely long facets}

\begin{figure}
    \centering
    \includegraphics{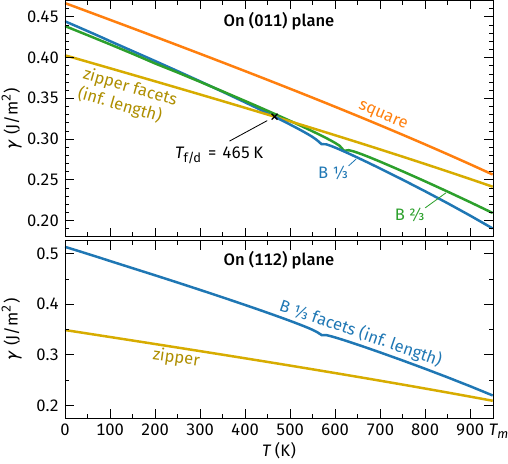}
    \caption{GB excess free energy as a function of temperature for
      the discovered GB phases. For the (011) plane, the temperature
      $T_\text{f/d}$ for the transition between zipper facets and a
      flat B phase is marked. The excess free energy of a faceted B
      phase on the (112) plane is so high that zipper stays stable up
      to the melting point $T_m$. For the B structures, there is a
      small dip in the data around \SI{600}{K}. This is an unphysical
      numerical issue with the potential, most likely a bond length
      being stretched beyond the potential's cutoff radius due to
      thermal expansion. Here, classical free energies are presented,
      which are the relevant values for classical MD
      simulations. Supplemental
      Fig.~\ref*{fig:suppl:free-energy-class-qm} shows that the
      inclusion of quantum-mechanical effects only significantly
      affects the excess free energy below around \SI{100}{K}.}
    \label{fig:free-energy}
\end{figure}

The thermodynamic (meta-)stability \footnote{Note that GBs are never
  thermodynamically stable. A system in thermodynamic equilibrium
  would only contain point defects. However, if we take the presence
  of a GB as a given, we can still formulate the relative stability of
  different GB phases using standard thermodynamics.} of GB phases can
be described by their excess free energy $\gamma$, which we calculated
using the QHA \cite{Foiles1994, Freitas2018}. To compare the GB energy
of faceted GBs to non-faceted GBs, it needs to be renormalized: the GB
area is $1/\cos\phi$ times larger in a GB faceted into segments
inclined by $\phi$ towards the average GB plane. Additionally, the
junctions between facets each contribute a defect energy
$E_\text{core}/t$, where $t$ is the length of the junction (here along
the tilt axis direction). Due to the GB excess stress and potentially
due to the dislocation character of the facet junction, elastic
interactions have to be included, too \cite{Hamilton2003}. This
results in \cite{Hamilton2003, Brink2024}
\begin{equation}
  \label{eq:gamma-facets}
  \gamma = \frac{\gamma_\text{facet}(T)}{\cos\phi} + \frac{E_\text{core}}{tL} + K\frac{\ln(L/\delta_0)}{L},
\end{equation}
where $L$ is the facet length and the last term represents the elastic
interactions with $K$ being a constant and $\delta_0$ being the
effective core size of the junction. For simplicity, we will assume
that the junction energies are independent of temperature. For
infinite facet lengths, only the first term remains nonzero, which is
also the lowest possible GB energy of a faceted $\Sigma3$
$[11\overline{1}]$ tilt GB \cite{Brink2024}.

Figure~\ref{fig:free-energy} shows the results of QHA calculations for
infinite facet lengths. A GB with a (011) plane would thus facet into
zipper segments at low temperatures and defacet at high temperatures
by transitioning into the B phase. A GB with a (112) plane, however,
would favor the zipper structure up to the melting point. This is
because the zipper phase is generally a low-energy GB phase, which is
only made unfavorable on the (011) plane by the increased GB area due
to faceting. The square phase does not play a role, due to its high
energy.

\subsection{Transformation kinetics}

\begin{figure}[b]
    \centering
    \includegraphics{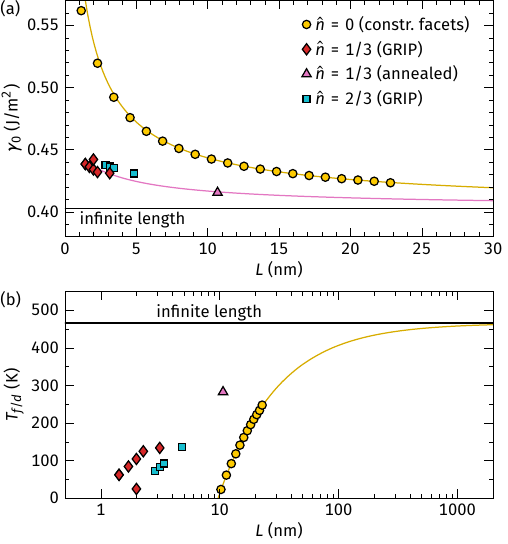}
    \caption{Influence of the finite length of \{112\} facets located
      on an average (011) GB plane. (a) Ground state energy $\gamma_0$
      of faceted zipper GBs as a function of facet length
      $L$. Differences in energy, especially at small $L$, are the
      result of different facet junction energies as a function of
      $\hat{n}$. The lines represent a fit of
      Eq.~\ref{eq:gamma-facets} to the data. Data points for
      $\hat{n}=0$ were constructed by joining zipper facets as
      described in Ref.~\cite{Brink2024}. For $\hat{n} = 1/3$ and
      $2/3$ the structures were obtained with \textsc{grip}. We
      replicated one sample with $\hat{n}=1/3$ along the $y$ direction
      and annealed it to obtain longer facet lengths. (b) By adding
      the facet junction energies to the excess free energies from
      Fig.~\ref{fig:free-energy}(a), we can estimate the change of the
      transition temperature $T_\text{f/d}$ for different facet
      lengths.}
    \label{fig:facet-length}
\end{figure}

In reality, when facets nucleate from a flat boundary, they need to
grow from short to long facet lengths. We therefore constructed
\{112\}-faceted GBs on the average (011) plane with varying lengths
and $\hat{n} = 0$ as described in Ref.~\cite{Brink2024}. We also
performed \textsc{grip} structure searches with larger supercells to
find faceted GBs with $\hat{n} \neq
0$. Figure~\ref{fig:facet-length}(a) shows the GB energies as a
function of facet length $L$. Equation~\ref{eq:gamma-facets} fits well
to the data (lines in the graph). The structures with $\hat{n} \neq 0$
differ by the type of their facet junction, but have the same facet
structure (Supplemental Fig.~\ref*{fig:suppl:zippers}), leading to
large GB energy differences at short facet lengths. This difference
becomes negligible at large facet lengths. We also estimated the
change of the faceting/defaceting transition temperature
$T_\text{f/d}$ in Fig.~\ref{fig:facet-length}(b), if the facet length
were constrained (e.g., due to a finite size simulation cell).
Transversely, this can be interpreted as a critical nucleus size as a
function of temperature.

Additionally, the zipper facets themselves exhibit values of
$\hat{n} = 0$, while $\hat{n} = 1/3$ or $2/3$ for the different B
structures. Consequently, the growth of faceted structures from the B
phase---or vice versa---requires either diffusion or leaves behind
defects in the GB. As such, we cannot take the theoretical value of
$T_\text{f/d}$ as a given and the actual kinetics in a real sample
will depend on the local point defect concentrations in and near the
GB.

\begin{figure}[t!]
    \centering
    \includegraphics{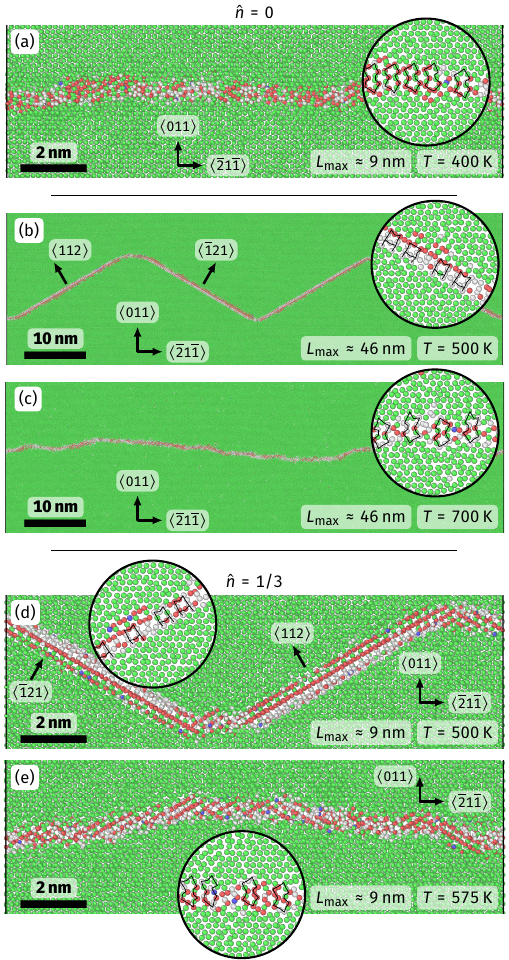}
    \caption{Snapshots from MD simulations of $\Sigma3$
      $[11\overline{1}]$ (011) GBs. (a) A GB with $\hat{n} = 0$ does
      not exhibit any faceting. Here, the simulation cell size admits
      a maximum facet length of $L_\text{max} = \SI{9}{nm}$ and
      therefore the facet junction energy is too high (see also
      Fig~\ref{fig:facet-length}). (b)--(c) When allowing facet
      lengths up to \SI{46}{nm}, a transition can be
      observed. (d)--(e) For $\hat{n} = 1/3$, the facet junction
      energies are lower. We observe faceting up to \SI{500}{K} (d),
      while a rough but flat GB occurs at \SI{575}{K} and above
      (e). In between, we observe a mix. The round insets show a zoom
      onto a slice with a thickness of 3 atomic layers, which allows
      to indentify the atomic motifs of the GB. We find zipper
      structures in the faceted GBs and B structures in the flat
      GBs. Indicated crystal directions refer to the lower
      crystallite. Equivalent crystal directions according to the
      misorientation in the upper crystallite not shown.}
    \label{fig:MD-sim-zhakhovskii}
\end{figure}

We nevertheless verified our results using various MD simulations. For
this, we constructed bicrystals with an average (011) GB plane, a
given $\hat{n}$ value, and a random GB structure and annealed them at
$T = 400$, 500, 600, and \SI{700}{K} for \SI{30}{ns}.

For $\hat{n} = 0$, we started with a periodic simulation cell with a
size in $y$ direction that admits a maximum facet length of
approximately \SI{9}{nm} [Fig.~\ref{fig:MD-sim-zhakhovskii}(a)]. Our
previous results predict that a faceted GB with those facet lengths
could never be stable [Fig.~\ref{fig:facet-length}(b)] and the MD
simulations support this: from 400 to \SI{700}{K}, we obtain flat GBs
with defective B structures. Larger simulations show sharply faceted
boundaries up to \SI{500}{K} and flat boundaries at \SI{700}{K}, with
a mix in between [Fig.~\ref{fig:MD-sim-zhakhovskii}(b)--(c)]. The
faceted GBs contain zipper structures, the flat ones defective B
structures (circular insets). These results can be reproduced both by
starting from an already faceted GB and from a flat GB. The transition
temperature significantly exceeds the prediction by the QHA
(Fig.~\ref{fig:free-energy}), but by inspecting
Fig.~\ref{fig:GRIP-zhakhovskii}, we can see that \textsc{grip} only
found high-energy GBs at $\hat{n} = 0$, meaning that the B phase is
much less stable there. A transition between $\hat{n}$ values would
require open surfaces \cite{Frolov2013} and much longer MD simulation
timescales than feasible for these system sizes.

For $\hat{n} = 1/3$ [Fig.~\ref{fig:MD-sim-zhakhovskii}(d)--(e)], the
facet junction energy is lower, and smaller facets can occur. We found
a transition temperature between 500 and \SI{575}{K}. This is still
higher than predicted by the QHA. However, the B structures are not
pristine, but the GBs are very rough. It appears that this
finite-temperature effect, which was not considered in the
\textsc{grip} structure search, raises the free energy of the B
structures and thereby the transition temperature. The free energy
differences of the GBs are also quite small and any numerical errors
can lead to quite large deviations in predicted transition
temperature. Qualitatively, the MD simulations support our theory,
especially since we find clear zipper structures in the faceted GBs
and (defective) B structures in the flat GBs (see circular insets in
Fig.~\ref{fig:MD-sim-zhakhovskii}).

\vfill%

\section{Discussion in the context of prior observations reported
  in the literature}

The original observation \cite{Hsieh1989} placed the
faceting/defaceting transition at roughly \SI{500}{K}, which fits well
with the free energy prediction and MD simulations given the expected
uncertainty of the transition temperature as discussed before. The
good agreement obtained with the Zha09 potential is due to the
accurate reproduction of the experimental melting point and of the GB
energies (see Appendix~\ref{sec:app:eval-pot}). Prior MD simulations
have qualitatively similar results as the original experiments and our
results, but did not investigate the transition as a GB phase
transition \cite{Wu2009}. We also found that the potentials used in
previous simulation studies are flawed (see
Appendix~\ref{sec:app:eval-pot}). Here, we therefore shed light on the
faceting/defaceting transition as a non-congruent GB phase transition.

In contrast, a study on epitaxial growth of Al films on alumina did
not reveal any defaceting upon annealing, but instead simply reported
the evolution of a maze structure of $\Sigma3$ $[11\overline{1}]$
\{112\} GBs \cite{Dehm2002}. Figure~\ref{fig:free-energy}(b) explains
this: GBs on \{112\} planes have no driving force to facet at all. The
zipper GB phase, which is native to this plane, has such low energy
that it will not transition into anything else. Consequently, any
\{112\} GB would be stable. The present study only considers the case
of a bicrystal with an average \{011\} GB plane. In a real polycrystal
the transition to \{112\} facets might not be reversible if the
microstructure reaches a deep energy minimum for its GB network. This
is consistent with the equilibrium shape of a polyhedral grain in an
infinite matrix: The interface stiffness tensor of Ni $\Sigma3$
$[11\overline{1}]$ boundaries shows that \{112\} facets are stable
compared to \{011\} facets \cite{Abdeljawad2018}. The same
considerations apply for Al---given the lower GB energy of \{112\}
facets compared to \{011\} facets---at least at low temperatures.

On the theory side, some previous models explain the
faceting/defacting transition as a faceting refinement (or roughening)
transition, where the configurational entropy of small facets
overcomes their energy cost at the critical temperature
\cite{Daruka2004, Qiu2023}. At low temperatures the facets are long
and clearly defined, at high temperatures the facets are small and
fluctuate. Direct atomistic simulations of this phenomenon were not
provided. Here, we instead clearly observe the structural change
between zipper facets and the B phase in MD simulations.

\section{Conclusion}

Experiments showed that $\Sigma3$ $[11\overline{1}]$ (011) tilt GBs in
Al are sharply faceted at lower temperatures, while becoming flat
above \SI{500}{K} \cite{Hsieh1989}. This phenomenon was investigated
and modeled in the intervening years \cite{Daruka2004, Wu2009,
  Straumal2016, Qiu2023}, but the exact nature of the transition
remained unclear. Here, we use atomistic computer simulations to show
that this faceting/defaceting transition is in fact a non-congruent,
first order GB phase transition. At low temperatures, the facets
consist of the low-energy zipper GB phase. At high temperatures, the
flat B phase becomes stable and the facets disappear. The faceting and
defaceting are therefore controlled by the thermodynamic stability of
their different atomic-scale structures.

We can relate this transition of the $\Sigma3$ GB
($\theta = \ang{60}$) to the congruent GB phase transitions in related
boundaries with lower misorientation angles ($\theta < \ang{60}$): The
B phase of the present work also appears as a motif in the pearl GB
phase of the $\Sigma37$c boundary \cite{Langenohl2022}. The zipper
facets are energetically driven to shrink to nanoscale size for
$\theta < \ang{60}$ and form the domino phase. That congruent
pearl-to-domino transition becomes a non-congruent faceting/defacting
transition for the $\Sigma3$ GB.

\newcommand{\h}[1]{\multicolumn{1}{c}{#1}}
\begin{table*}
    \centering
    \caption{Material properties of Al computed with the
      (M)EAM potentials compared to literature values
      (ref.).
      The reference values for the experimental ground-state energies
      $E_0^\text{fcc}$ are from Ref.~\cite{Kittel2005}, while
      the experimental lattice constant $a_0^\text{fcc}$, elastic
      constants $c_{ij}$, and the melting point $T_m$ are from
      Ref.~\cite{CRCHandbook2022}. The energy difference $E_\text{fcc$\rightarrow$hcp}$
      between hcp and
      fcc crystal structures was calculated in the present work using DFT (we used primitive unit cells and a $21\times21\times21$ $k$-point grid, but otherwise the same parameters as described in Sec.~\ref{sec:methods:dft}).
      The vacancy formation energy $E_{f,\text{vac}}$ is
      also an experimental value \cite{Ullmaier1991}. The surface energy
      $\gamma_{(111)}$ is from DFT simulations \cite{Vitos1998}. We
      used the literature review in Ref.~\cite{Bernstein2004} to
      collect data for values of the experimental stacking-fault energy
      $\gamma_\text{SF}$ and DFT values for the unstable
      stacking-fault energy $\gamma_\text{USF}$. The maximum shear
      stress $\tau_\text{SF}$ along the generalized stacking-fault
      curve is from DFT calculations \cite{Ogata2002}.}
    \label{tab:bulk-prop}
    \newcolumntype{d}[1]{D{.}{.}{#1}}
    \begin{ruledtabular}
    \begin{tabular}{ld{2.3}d{1.3}d{2.3}d{3.0}d{2.0}d{2.0}d{4.0}d{1.2}d{1.3}d{3.1}d{3.1}d{1.1}}
        & \h{$E_0^\text{fcc}$} & \h{$a_0^\text{fcc}$} & \h{$E_\text{fcc$\rightarrow$hcp}$} & \h{$c_{11}$} & \h{$c_{12}$} & \h{$c_{44}$} & \h{$T_m$} & \h{$E_{f,\text{vac}}$} & \h{$\gamma_{(111)}$} & \h{$\gamma_\text{SF}$} & \h{$\gamma_\text{USF}$} & \h{$\tau_\text{SF}$} \\[2pt]
        & \h{(eV/atom)}        & \h{(\AA)}            & \h{(eV/atom)}        & \h{(GPa)}    & \h{(GPa)}    & \h{(GPa)}    & \h{(K)}             & \h{(eV)}               & \h{(J/m$^2$)}        & \h{(mJ/m$^2$)}         & \h{(mJ/m$^2$)}          & \h{(GPa)}            \\[2pt]
    \colrule
    ref.  & -3.39  & 4.050 &  0.03  & 107 & 60 & 28 &  933 & 0.67 & 1.199 & \h{135--200} & \h{175--224} & 2.8 \\
    \colrule
    Zha09 & -3.361 & 4.032 &  0.003 & 105 & 70 & 44 &  951 & 0.74 & 0.733 &  15.9 & 107.2 & 1.9 \\
    Vot86 & -3.360 & 4.050 &  0.014 & 107 & 65 & 32 &  602 & 0.63 & 0.820 &  75.5 &  93.1 & 1.5 \\
    Jac87 & -3.388 & 3.988 & -0.001 & 111 & 85 & 46 &\h{--}& 1.16 & 0.908 &  -6.0 &  88.1 & 1.8 \\
    Mis99 & -3.360 & 4.050 &  0.028 & 114 & 62 & 32 & 1042 & 0.68 & 0.870 & 145.5 & 167.3 & 2.3 \\
    Stu00 & -3.390 & 4.050 &  0.005 &  95 & 68 & 42 &  933 & 0.89 & 0.618 &  21.3 &  99.2 & 1.8 \\
    Zop03 & -3.360 & 4.050 &  0.022 & 117 & 60 & 32 &  871 & 0.71 & 0.601 & 113.8 & 149.7 & 2.3 \\
    Liu04 & -3.360 & 4.032 &  0.024 & 119 & 63 & 33 &  884 & 0.68 & 0.912 & 129.5 & 163.4 & 2.3 \\
    Zho04 & -3.580 & 4.050 &  0.001 & 105 & 60 & 28 &  577 & 0.67 & 0.909 &\h{--} &\h{--} &\h{--}\\
    Men08 & -3.411 & 4.045 &  0.028 & 105 & 59 & 31 &  926 & 0.66 & 0.428 & 127.1 & 219.7 & 3.7 \\
    Win09 & -2.646 & 4.025 &  0.030 & 114 & 62 & 31 &  846 & 0.66 & 0.876 & 140.9 & 178.9 & 2.6 \\
    Pas15 & -3.360 & 4.050 &  0.040 & 114 & 62 & 45 &  944 & 0.67 & 0.718 & 186.5 & 305.0 & 4.4 \\
    \end{tabular}
    \end{ruledtabular}
\end{table*}

\section{Data availability}

The data that support the findings of this article are openly
available \cite{zenodo}.

\section{Acknowledgments}

The authors thank Saba Saood for providing insights into her
experimental work on Al GBs. This work was supported by the National
Research Foundation of Korea (NRF) funded by the Korea government
(Ministry of Science and ICT) (No.\
RS-2023-00254343). T.B. acknowledges funding from the European
Research Council (ERC) under the European Union's Horizon 2020
research and innovation program (Grant agreement No. 787446;
GB-CORRELATE).

Y.C. conducted the bulk of the simulations, while T.B. was responsible
for some supplemental simulations. Y.C. and T.B. analyzed the data and
wrote the initial manuscript draft. T.B. conceptualized and supervised
the project.

\raggedbottom   

\appendix

\section{Evaluation of empirical potentials}
\label{sec:app:eval-pot}

We considered a range of classical empirical potentials from the
literature and the NIST Interatomic Potentials Repository \cite{ipr}
for our simulations. We used several EAM potentials, designated as
Vot86 \cite{Voter1986}, Jac87 \cite{Jacobsen1996}, Mis99
\cite{Mishin1999}, Stu00 \cite{Sturgeon2000}, Zop03 \cite{Zope2003},
Liu04 \cite{Liu2004a}, Zho04 \cite{Zhou2004}, Men08
\cite{Mendelev2008}, Win09 \cite{Winey2009}, and Zha09
\cite{Zhakhovskii2009}, as well as an MEAM potential, Pas15
\cite{Pascuet2015}. We reconstructed the Vot86 potential based on the
data in the original publications (the notebook and resulting
potential file is available in the supporting data \cite{zenodo}). The
Mis99 potential is the modified version described in
Refs.~\cite{Brink2023, Brink2023zenodo}, which should give the same
results as the original potential, apart from reducing numerical noise
especially in the second derivatives required for phonon
calculations. The parametrization for Jac87 was extracted from
Ref.~\cite{Norskov2019kim}, which is part of the OpenKIM project
\cite{Tadmor2011, Elliott2011}. All other potentials were obtained
from the NIST Interatomic Potentials Repository \cite{ipr}.

We then proceeded to evaluate the accuracy of these potentials against
reference data from experiment or DFT. Bulk properties for fcc Al and
the cohesive energy for hcp Al have been calculated as described in
Ref.~\cite{Brink2023} and the results are listed in
Table~\ref{tab:bulk-prop}. Cohesive energies, lattice constants, and
elastic constants are reasonable for all potentials, except for Jac87
in which hcp is the stable phase and except for Win09 whose cohesive
energy is too low. We therefore did not consider Jac87 in further
simulations. The energy difference between fcc and hcp is often
underestimated, as discussed below in more detail in the context of
the stacking-fault energy. The $c_{44}$ elastic constant is often
slightly overestimated. Melting point calculations were taken from
Ref.~\cite{Zhu2021} where available, otherwise they were computed with
the software from that publication. Vot86 and Zho04 significantly
underestimate the melting point, while the closest matches are Stu00,
Men08, Zha09, and Pas15. Vacancy formation energies are once again in
a reasonable range for all potentials, except Jac87. The (111) surface
energies are underestimated by all potentials, which is a known issue
with EAM-type potentials. When calculating the stacking-fault
energies, the Zho04 potential yielded different results when
constructing the stacking fault compared to producing it by gradual
displacement of two halves of a perfect crystal. The latter is
performed to obtain the unstable stacking-fault energy via a
generalized stacking-fault energy curve \cite{Vitek1966, Vitek1968,
  Zimmerman2000, Cai2002}. Both methods must yield the same result,
which points to numerical issues with the Zho04 potential, where
slightly different defect configurations have significantly different
energies. We therefore also excluded the Zho04 potential from further
consideration. The Mis99, Liu04, Men08, and Win09 potentials yield
(unstable) stacking-fault energies within the ranges reported in the
literature. These potentials should therefore be preferred when
simulating plasticity. Our potential of choice---Zha09---is poorly
reproducing the stacking-fault energies. As outlined below, it is
however an excellent match for the GB excess properties required in
this paper.

\begin{figure}
    \centering
    \includegraphics{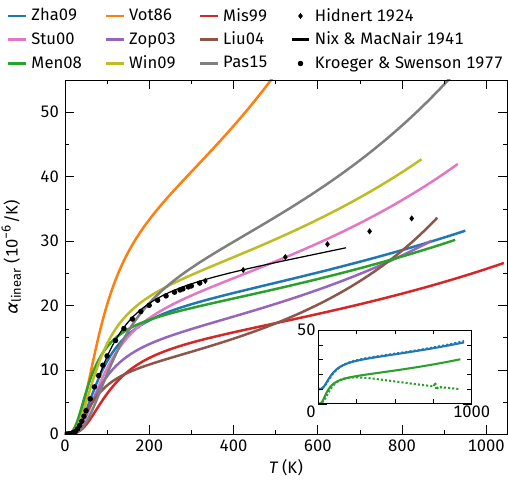}
    \caption{Linear thermal expansion coefficients. Black points are
      experimental data from Hidnert \cite{Hidnert1924}, as well as
      Kroeger and Swenson \cite{Kroeger1977}. The black line is a fit
      to the Gr\"uneisen equation of state from Nix and MacNair
      \cite{Nix1941}. Results for the empirical potentials were
      obtained by equilibrating with a QTB at \SI{50}{K} temperature
      intervals and subsequently fitting the Gr\"uneisen equation of
      state to the data. The plots end at the melting point. The inset
      shows a comparison of this method to thermal expansion computed
      with the QHA for Zha09 and Men08 (dotted lines). Due to
      numerical problems with other potentials, only Zha09 has a match
      between methods (see also Supplemental
      Fig.~\ref*{fig:suppl:latconst}).}
    \label{fig:thermal-expansion}
\end{figure}

We also require a reasonable description of free energies as a
function of temperature. Figure~\ref{fig:thermal-expansion} shows the
linear thermal expansion coefficients compared with literature
values. The combination of thermal expansion coefficient and melting
point is a good indicator of the quality of the free energy as a
function of temperature and the values are reliably accessible
experimentally. In order to include quantum effects into the otherwise
classical MD simulations, we used a quantum thermal bath (QTB)
\cite{Dammak2009, Barrat2011} instead of the typical classical
thermostats. We then fitted a Gr\"uneisen equation of state
\cite{Grueneisen1926, Nix1941} to the data and show it in
Fig.~\ref{fig:thermal-expansion} alongside literature data
\cite{Hidnert1924, Nix1941, Kroeger1977}. Note that the QTB approach
is approximate \cite{Dammak2009comment, Dammak2009reply, Barrat2011},
but we ensured that the results coincide with the thermal expansion
with classical thermostats at high temperatures (see Supplemental
Fig.~\ref*{fig:suppl:latconst}). Furthermore, we also calculated the
thermal expansion using the QHA with \textsc{phonopy} \cite{Togo2023,
  Togo2023a}. We found, however, that the QHA approach is unreliable
with most potentials; only Zha09 coincides at all temperatures (see
inset of Fig.~\ref{fig:thermal-expansion} and Supplemental
Fig.~\ref*{fig:suppl:latconst} for a more detailed discussion). We
conclude that only Zha09 has both reasonable values for thermal
expansion and melting point, as well as reliable QHA results across
the temperature range.

\begin{figure}
    \centering
    \includegraphics{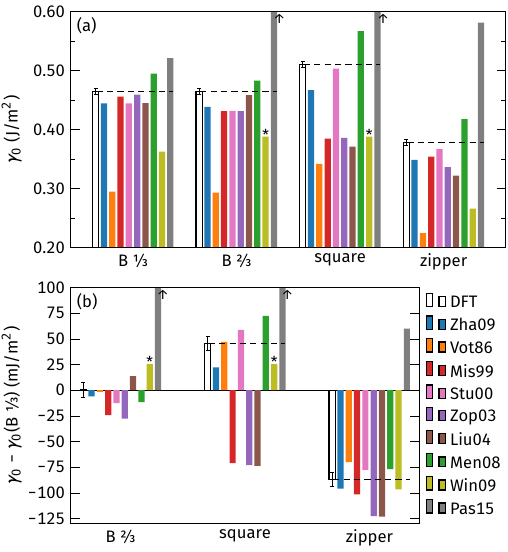}
    \caption{(a) Comparison between GB energies calculated with DFT and
      with empirical potentials. The DFT error bars are conservatively
      estimated to be $\SI{+-5}{mJ/m^2} = \SI{+-0.3}{meV/\angstrom^2}$
      from variations in the GB energy with system size during
      convergence studies (Supplemental
      Figs.~\ref*{fig:suppl:dft-B1}--\ref*{fig:suppl:dft-square}). The
      B~\sfrac{2}{3} structure minimized with the Win09 potential
      (marked by an asterisk) transforms into the square
      structure. (b) GB energy difference compared to B~\sfrac{1}{3}.}
    \label{fig:verification:gamma0}
\end{figure}

We next move on to discuss the suitability of the potentials for
simulating GBs. For this, we first used DFT calculations to establish
the excess GB properties for the B~\sfrac{1}{3}, B~\sfrac{2}{3},
square, and zipper GB phases. We used the calculations from
Ref.~\cite{Saood2023} for zipper, and calculated the missing values
here (Supplemental
Figs.~\ref*{fig:suppl:dft-fcc}--\ref*{fig:suppl:dft-square}). We
obtained the GB energy $\gamma_0$ and compare it with the results
obtained using the interatomic potentials in
Fig.~\ref{fig:verification:gamma0}. We find that the Mis99, Zop03, and
Liu04 potentials predict that the square GB phase is more stable than
the B phase, which is contradicted by DFT. Furthermore, when using the
Win09 potential, the B \sfrac{2}{3} structure becomes unstable and
transforms into the square structure. The GB energies using the Pas15
potential are generally too high and the zipper phase's energy is much
too high compared to the B \sfrac{1}{3} structure. We therefore
exclude these potentials from further consideration for our GBs, as
they would not be able to accurately predict the GB phase
transition. Additionally, we ran \textsc{grip} structure searches for
all potentials (Supplemental Fig.~\ref*{fig:suppl:grip-allpots}). The
results confirm the calculations performed on the B~\sfrac{1}{3},
B~\sfrac{2}{3}, square, and zipper structures.

\begin{figure}
    \centering
    \includegraphics{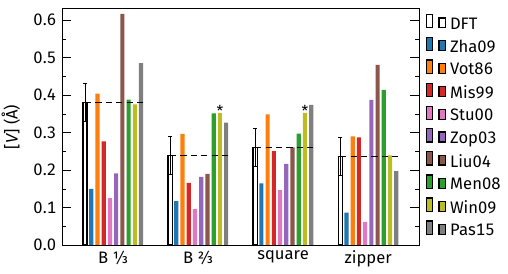}
    \caption{Comparison between excess volumes calculated with DFT and
      with empirical potentials. The DFT error bars are conservatively
      estimated to be \SI{+-0.05}{\angstrom} from variations in the GB
      energy with system size during convergence studies (Supplemental
      Figs.~\ref*{fig:suppl:dft-B1}--\ref*{fig:suppl:dft-square}). The
      B~\sfrac{2}{3} structure minimized with the Win09 potential
      (marked by an asterisk) transforms into the square structure.}
    \label{fig:verification:V}
\end{figure}

Out of the remaining potentials, the Vot86 and Men08 potentials
reproduce the excess volume better than the Zha09 and Stu00 potentials
(Fig.~\ref{fig:verification:V}). We believe this to be less important
than accurate energies and thermal expansion for our present
investigation and therefore do not dismiss the Zha09 and Stu00
potentials here.

\begin{figure}
    \centering
    \includegraphics{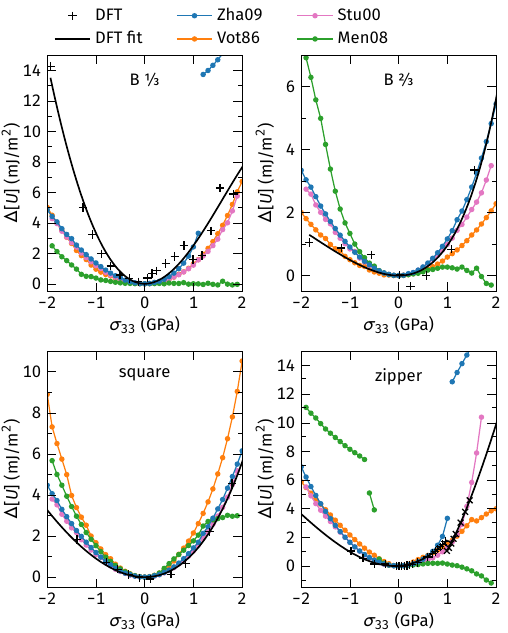}
    \caption{Change of excess energy with applied stress
      $\sigma_{33}$. For DFT, the data points represent individual
      calculations and the lines are polynomial fits to the data.}
    \label{fig:verification:sigma33:U}
\end{figure}
\begin{figure}
    \centering
    \includegraphics{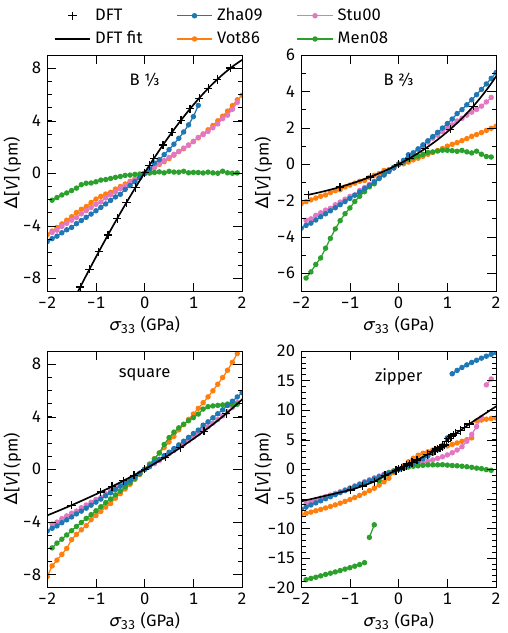}
    \caption{Change of excess volume with applied stress
      $\sigma_{33}$. For DFT, the data points represent individual
      calculations and the lines are polynomial fits to the data.}
    \label{fig:verification:sigma33:V}
\end{figure}

Instead, we investigated the GBs under different applied stresses
$\sigma_{33}$. The change of excess potential energy with stress is
presented in Fig.~\ref{fig:verification:sigma33:U}. None of the
potentials perfectly match the GB's stiffness calculated with
DFT. However, the Zha09 potential is a close match under tension,
although it can be too soft or too stiff under compression. The Stu00
potential is of comparable quality and the Vot86 potential is
qualitatively correct, but matches the DFT curves less well. The Men08
potential seems to exhibit softening/instability under tension. For
the zipper phase, a transition between microstates under stress or
strain has been observed before \cite{Saood2023} and manifests here as
a discontinuity in the energy curve. Such a transition between
microstates is simply a sudden, but slight rearrangement of a GB atom
in a shallow double-well potential. The Zha09 potential reproduces
this transition at \SI{1}{GPa} accurately, although the change in
energy is much too high. The Vot86 and Stu00 potential predict this
transition at the wrong stress and the Men08 potential even exhibits
the transition under compression instead of tension. A similar
microstate transition occurs with the Zha09 potential in the
B~\sfrac{1}{3} structure, but DFT calculations are inconclusive: some
scatter is observed in the data, but no clear trends could be found.
All of these findings are qualitatively matching to the evolution of
excess volume under stress (Fig.~\ref{fig:verification:sigma33:V}).

\begin{figure}[b!]
    \centering
    \includegraphics{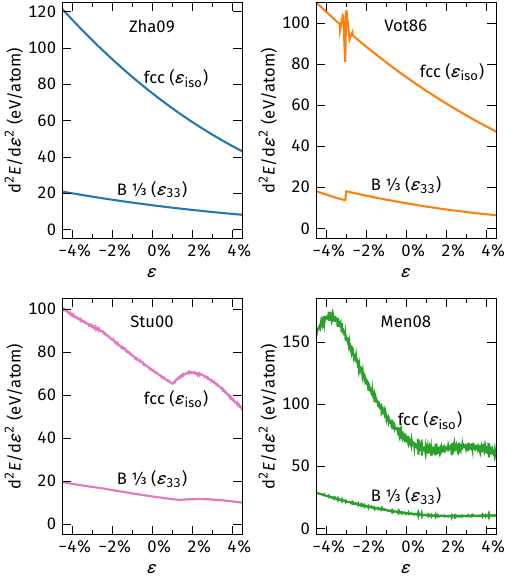}
    \vspace{-2\baselineskip}%
    \caption{Second derivative of the potential energy with strain,
      computed using finite differences. For each potential, both an
      fcc unit cell and a B~\sfrac{1}{3} GB were strained; fcc
      hydrostatically and the GB along the $z$ direction. These plots
      reveal numerical problems with cutoffs, noise, and fluctuating
      curvature of the empirical potentials.}
    \label{fig:verification:strain-for-energy-derivatives}
\end{figure}

Finally, numerical issues can occur when derivatives of the potential
energy are required, such as in the case of QHA, where the force
constant matrix contains the second derivatives of the potential
energy. This is because numerical derivates amplify any noise or
discontinuities present in the interatomic potential. We tested all
potentials by starting once from an fcc unit cell and once from a
B~\sfrac{1}{3} GB and then straining the simulation box
(Fig.~\ref{fig:verification:strain-for-energy-derivatives} and
Supplemental Figs.~\ref*{fig:suppl:derivatives-a} and
\ref*{fig:suppl:derivatives-b}). We used fine-grained scaling factors,
recorded the potential energy at each step (without additional
relaxation of the atomic positions for computational efficiency), and
calculated the derivatives using the finite difference method. Of the
potentials under consideration, only the Zha09 potential is smooth,
which is the expected result. The Vot86 potential exhibits some noise
at a specific strain, which is most likely related to atoms leaving
the cutoff radius. The Stu00 potential contains slight numerical
noise, as well as a discontinuity in the second derivative. The Men08
potential is somewhat more noisy. Supplemental
Figs.~\ref*{fig:suppl:derivatives-a} and
\ref*{fig:suppl:derivatives-b} contain data for all potentials and
also explain some of the previously observed shortcomings of
potentials we excluded from consideration.

In summary, we chose the Zha09 potential for its accurate reproduction
of GB energies and thermal expansion coefficients. It is also smooth
and the thermal expansion computed using QHA coincides with the QTB
method and matches well to the experiments. We would recommend this
potential for GB (free) energy calculations and GB structure search as
performed in the present paper. The main shortcoming we observed is
the underestimation of the stacking-fault energy, which makes Zha09
unsuitable for simulations of plasticity. For MD simulations of
plasiticity that also include GBs, we would recommend the Men08
potential, even though it has some numerical issues.

\end{document}